\documentclass[prl,twocolumn,showpacs,preprintnumbers,amsmath,amssymb]{revtex4}
\usepackage{dcolumn}
\usepackage{bm}
\usepackage{graphicx}
\usepackage{bm}
\begin{document}
\title{Dilution of 2D antiferromagnetism by Mn site substitution in  La${_1}$Sr${_2}$Mn${_{2-x}}$Al${_x}$O${_7}$}
\author{Sunil Nair and A. Banerjee}
\affiliation{Inter University Consortium for D.A.E. Faclilites,\\ University Campus, Khandwa Road, Indore, 452 017, INDIA.}
\date{\today}
\begin{abstract}
We report the effect of Al substitution on the Mn site of the bilayered half doped manganite La${_1}$Sr${_2}$Mn${_2}$O${_7}$. This substitution dilutes the magnetically active Mn-O-Mn network without introducing an appreciable distortion in the lattice and ionic considerations leads to a predominant reduction of Mn${^{4+}}$ with increasing Al. The rate of fall of the long range antiferromagnetic transition temperature as a function of substitution is seen to match well with established quasi 2D Heisenberg systems indicating that the nature of magnetic interactions in this quasi 2-D system is of the short range Heisenberg type. The magnetic contribution of the specific heat estimated using Fisher's relation could be fitted with a function incorporating the presence of a gapped Fermi surface appropriate for this type of systems. The resistivity was seen to increase as a function of substitution due to the weakening of the double exchange within the ferromagnetic layers of these A type of antiferromagnets and justifiably, in the paramagnetic region the data could be fitted to Mott's equation for the variable range hopping of polarons in 2 dimensions. 
\end{abstract}
\maketitle
Mn based Ruddlesden-Popper\cite{RP} compounds (LaSr)${_{n+1}}$Mn${_n}$O${_{3n+1}}$ are known to have a layered structure in which \textit{n}-MnO${_3}$ blocks are seperated by an additional (La${_{1-x}}$Sr${_x}$)O layer. The perovskite structure corresponds to the n $=\infty$ and the K${_2}$NiF${_4}$ structure to n $= 1$ members of this series. The bilayered (n $= 2$) manganites La${_{2-2x}}$Sr${_{1+2x}}$Mn${_2}$O${_7}$ have been extensively investigated ever since the x$ = 40\%$ member of this series showed collosal magnetoresistance\cite{mori}, albeit at lower temperatures.  The structural consequence of moving from the perovskite (n$ =\infty$) to the bilayer (n $=2$) is the introduction of a 2 dimensional character to the system.  The number of next neighbour Mn cations around each transition metal site reduces from 6 to 5, thus providing an anisotropic reduction in the one electron (e$_g$) bandwidth, as a consequence of which the electronic and transport properties are interestingly modified.

One of the most intruiging phenomenon observed in perovskite manganites is the real space ordering of Mn${^{3+}}$ and Mn${^{4+}}$ ions. This \emph{charge ordering}\cite{rao}is frequently observed at commensurate compositions, provided the bandwidth (W) is not large. In the vicinity of 50$\%$ hole doping, a variety of perovskite manganites show the so-called CE type of spin ordering, which is made up of zig-zag ferromagnetic chains ordered antiferromagnetically\cite{good1}.  In analogy with the 3D manganites, the bilayered manganite La${_1}$Sr${_2}$Mn${_2}$O${_7}$ is also expected to show similar spin/charge ordering considering the fact that it has an equal (50$\%$) of Mn${^{3+}}$ and Mn${^{4+}}$ ions.  Though initial diffraction experiments supported this point of view\cite{tokura1}, later results have indicated that the CE type of spin/charge ordering exists only at intermediate temperatures\cite{chat} and that its volume fraction is small as compared to the much more dominant A type of antiferromagnetic(AFM) state\cite{kubota}. 

Mn site substitution, besides introducing various interesting phases\cite{hardy} in the half doped perovskite manganites offers considerable insight into the nature of magnetism and transport in this class of materials.  Though this avenue of research has gained popularity in recent times, Mn site substitution in the layered analogues have not been adequately explored.  Here, we report the magnetic and transport properties of the bilayered manganite La${_1}$Sr${_2}$Mn${_2}$O${_7}$, with Al substituted in the Mn site.  Al was selected as a dopant on ionic considerations, as its ionic radii matches very well with that of Mn${^{4+}}$\cite{shan}. Thus one would expect minimal lattice distortion along with a preferential replacement of Mn${^{4+}}$. Most importantly, Al with an empty d shell is non magnetic, and hence a simple dilution of the magnetic lattice would be expected. 

Polycrystalline samples of La${_1}$Sr${_2}$Mn${_{2-x}}$Al${_x}$O${_7}$ ($0\leq x \leq 10\%$) have been prepared using the standard solid state ceramic route, with starting materials La${_2}$O${_3}$, SrCO${_3}$, MnO${_2}$ and Al${_2}$O${_3}$ of atleast 99.99$\%$ purity. The powdered samples are mixed and initially treated for 1000\raisebox{1ex}{\scriptsize o}C for 24 hours after which they are pelletised, reground and treated for 1250\raisebox{1ex}{\scriptsize o}C (24 hrs) and 1500\raisebox{1ex}{\scriptsize o}C (36hrs) with intermediate grindings. X ray diffraction is done using a Rigaku Rotaflex RTC 300RC powder diffractometer with Cu K$\alpha$ radiation. All the samples are seen to crystallise in the tetragonal (\textit{I4/mmm}) structure.  Rietveld profile refinement\cite{young} of powder XRD data is used to determine the structural parameters tabulated in Table.1. As is clearly seen, the variation in the structural parameters is $<$ 1$\%$ indicating that Al doping does not introduce any major structural distortion. The values of the mean Mn valence determined by using iodometric redox titrations (using sodium thiosulphate and pottasium iodide) unambiguously indicate a preferential replacement of Mn${^{4+}}$ with increasing Al substitution. The presence of stacking faults in the double layered structure is known to result in the formation of parasitic (LaSr)MnO${_3}$ and (LaSr)${_2}$MnO${_4}$ phases\cite{intergrowth}. A parasitic perovskite phase has been detected in our samples and (using Rietveld profile refinement) is estimated to make up $\approx$ 3$\%$ of the volume fraction.  However we have reasons to believe that the presence of this intergrowth does not adversely affect the nature of magnetism and transport in these samples at least in the temperature ranges of our interest. 

\begin{table}
\caption{\label{tab:table 1}Structural and fitting parameters determined from the Rietveld profile refinement of the powder XRD patterns for the series La${_{1}}$Sr${_{2}}$Mn${_{2-x}}$Al${_x}$O${_{7+\delta}}$.  Here O1 refers to the apical oxygen in the double perovskite slab, O2 is the equitorial oxygen which lies in the plane of the perovskite layer, and O3 is the apical oxygen in the rocksalt layer. The mean valence state of Mn was determined by redox iodometric titrations}
\begin{ruledtabular}
\begin{tabular}{cccccc}
Al (x) &$0 \%$ &$2.5 \%$  &$5 \%$ &$7.5 \%$ &$10 \%$\\
\hline
a($\AA$) &3.8707 &3.8690 &3.8652 &3.8618 &3.8544\\
c($\AA$) &19.9771 &19.9820 &19.9921 &19.9982 &20.0006\\
V($\AA{^3}$) &299.30 &299.11 &298.68 &298.24 &297.14\\
Mn-O1($\AA$)  &1.949 &1.958 &1.969 &1.976 &1.984\\
Mn-O2($\AA$)  &1.951 &1.949 &1.947 &1.946 &1.943\\
Mn-O3($\AA$)  &1.962 &1.966 &1.982 &1.998 &2.032\\
Mn-O2-Mn  &165.41\raisebox{1ex}{\scriptsize o} &165.51\raisebox{1ex}{\scriptsize o} &165.66\raisebox{1ex}{\scriptsize o} &165.82\raisebox{1ex}{\scriptsize o} &165.98\raisebox{1ex}{\scriptsize o}\\
Mn${^{3+}}$\%  &49.0 &48.5 &47.6 &46.7 &45.7\\
Mn${^{4+}}$\% &51.0 &49.0 &47.4 &45.8 &44.3\\
\end{tabular}
\end{ruledtabular}
\end{table}

Magnetic susceptibility measurements are done using a home made ac susceptometer\cite{ashna}. The parent compound La${_1}$Sr${_2}$Mn${_2}$O${_7}$ is known to order antiferromagnetically at $\approx$ 225K, and since Al substitution dilutes the magnetically active Mn-O-Mn network, one would expect the transition temperatures to drop to lower temperatures with increasing Al substitution. This is clearly seen in Fig.1, where the temperature dependence of the real part of ac-susceptibility for the whole series shows a rapid reduction of the Neel temperature (T${_N}$) as a function of increasing Al doping. 

This problem of site dilution has been extensively studied in the past in an attempt to understand the nature of percolation in magnetic systems, and the rate of fall in T${_N}$ is known to depend on the universality class to which the system belongs. For instance, according to the mean field theory the long range T${_N}$ decreases linearly with increasing non magnetic substitution and long range order dissapears only when all the magnetic ions are substituted by non magnetic ones. For quasi 2-D Ising systems (like K${_2}$Co${_2}$F${_4}$ \cite{breed}), this drop is known to be much more rapid, and long range order dissapears at a dopant concentration of x $\approx$ 40$\%$, which is the site percolation threshold for a 2D lattice. However for systems like La${_2}$Cu${_{1-x}}$Mg${_x}$O${_4}$\cite{swc} which are known to be 2D Heisenberg like, the fall in T${_N}$ is even more drastic, and long range order is expected to dissapear at x $\approx$ 20-25$\%$. A useful parameter for the comparison of various systems is the Initial suppression rate (R${_{IS}}$) defined as -{$\partial$[T${_N}$(x)/T${_N}$(0)]/$\partial$x}. For our system, we obtain R${_{IS}}$$\approx$ 2.7 which is close to the that of other 2D Heisenberg systems\cite{swc}. This is shown in Fig. 2 where the fall in T${_N}$ is shown as a function of substitution. Our data is seen to match very well with that of  La${_2}$Cu${_{1-x}}$Mg${_x}$O${_4}$, a well established 2D Heisenberg system. \emph{Thus our results clearly indicate that the nature of magnetic interactions in this bilayered series are of the short range 2D Heisenberg type}. However, rigorous measurements of the critical region can be made to reconfirm this observation. 

It has been proposed\cite{fisher} (and verified\cite{grif}) that in a paramagnetic-antiferromagnetic transition, the magnetic contribution to the specific heat (C${_{mag}}$) mirrors the behaviour of $\partial(\chi T)/\partial T$; \textit{ie} C${_{mag}} = A \partial(\chi T)/\partial T$ where A is a relatively slowly varying function of temperature. Fig. 3 shows the temperature dependence of $\partial(\chi T)/\partial T$ for the whole series. All the samples show a typical $\lambda$-type feature which is a signature of an antiferromagnetic transition. The temperature of the peak of this feature is marginally smaller than the temperature where the peak in susceptibility occurs, as is expected in layered systems, where the effect of short range (2-D) correlations are known to be more pronounced. The data close to the transition region for T$<$ T${_N}$ could be fitted to an equation \\\\
$\partial(\chi T)/\partial T \propto C{_{mag}}$ = $\alpha$  exp(-$\Delta/T$)\\

where the exp (-$\Delta$/T) term arises from the presence of a gapped Fermi surface\cite{maple}.  Good fits could be obtained for all the samples in the range $\approx$ 1/3(T/T${_N}$), and the magnitude of $\Delta$ was seen to reduce with increasing Al doping as is shown in the inset of Fig.3. The origin of the gap ($\Delta$) used in fitting our data is not easy to discern.  Possible mechanisms that can be considered include the effect of spin, charge or orbital order, a Coulomb gap, strong electron lattice coupling or even a simple splitting of the levels due to the Jahn Teller effect on the Mn${^{3+}}$ ions. The data below $\approx$ 1/3(T/T${_N}$) could be fitted incorporating an addition term (T${^3}$) arising due to the presence of antiferromagnetic spin waves (magnons)\cite{esr}. However, we have refrained from including it in our analysis, considering the fact that the constant of proprtionality (A) used in the Fisher's relation is a slowly varying function of T \textsl{in the vicinity} of the transition, and hence fitting the data far away from this region could result in considerably larger errors. 

Determination of the nature of electronic transport in hole doped manganites is a problem that continues to elude solutions\cite{sal}. Current consensus is that the conduction atleast in the paramagnetic regime occurs through the hopping of charged carriers localised in the form of polarons. These polarons could be \textit{dielectric} polarons where the electron bears with it a dilation of the MnO${_6}$ octahedron, or a \textit{Jahn-Teller} polaron due to the axial distortion of the octahedron. The paramagnetic state of the layered manganites have recieved much lesser attention, though early studies on La${_{1.2}}$Sr${_{1.8}}$Mn${_2}$O${_7}$ drastically demonstrated the effect of dimensionality, as the conduction along the \textsl{ab} plane indicated the presence of \textit{Zener-pair} polarons, whereas along the \textsl{c} axis adiabatic small polaronic behaviour was concluded\cite{zhou}. Recently, the conduction in the paramagnetic state of La${_1}$Sr${_2}$Mn${_2}$O${_7}$ was reported to be due to the variable range hopping of polarons in the presence of a Coulomb gap\cite{vrh}. In general, the Mott's Variable range hopping(VRH)\cite{mott} is described by $\rho$= $\rho{_0}$ exp [T${_0}$/T]${^p}$, where p = 1/(d+1) with 'd' being the dimensionality of the system. Mott's activation energy (T${_0}$) $\propto$ 1/\textit{N(E)}$\xi{^d}$, where \textit{N(E)}is the density of states at the Fermi level and $\xi$ is the localisation length. However, in our case the transport data in the paramagnetic region of all the samples is seen to have much better fit to Mott's equation for VRH in 2 dimensions(p = 1/3) than to a VRH in the presence of a Coulomb gap (p = 1/2) as is seen in Fig.4. The values of T${_0}$ were seen to decrease with increasing Al doping, implying an increase in the localisation length provided the density of states at the Fermi level does not change.  

The transport data in the region T $<$ T${_N}$ for all the samples is shown in Fig.5, where the resistivity is seen to increase with increasing Al doping. This could be understood to be due to the destabilisation of the A type of AFM state. This A type of AFM is known to occur in systems with a relatively large bandwidth\cite{kawano} and is of a 2-D character, where the e${_g}$ electrons in each \textit{ab} (ferromagnetic) plane are itenerant. Here the e${_g}$ electrons supposedly occupy the isotropic (and delocalised) \textit{d(x${^2}$-y${^2}$)} orbitals\cite{hiro}, in contrast to the CE-type of half doped AFMs, where the e${_g}$ electrons occupy the anisotropic (and localised) \textit{d(3x${^2}$-r${^2}$)} and \textit{d(3y${^2}$-r${^2}$)} orbitals alternately\cite{mura}. These ferromagnetic layers are then alligned antiferromagnetically, making up this anisotropic antiferromagnet. Al substitution randomly cuts the electronically and magnetically active Mn-O-Mn network and reduces the effective double exchange strength within these ferromagnetic layers, resulting in a net increase in the resistivity. This behaviour is at variance with that seen in a 3D perovskite charge ordered system Pr${_{0.5}}$Ca${_{0.5}}$MnO${_3}$with a CE type of AF ordering, where Al substitution was seen to reduce the value of the resistivity in the (weakened) charge ordered region\cite{sunil}.\emph{This clearly brings forward the contrasting effects non magnetic substitution has on half doped manganites with different (ie A or CE type) antiferromagnetic ground states}.

In summary, polycrystalline samples of Al substituted La${_1}$Sr${_2}$Mn${_2}$O${_7}$ were prepared to study the effect of non magnetic substitution in the half doped bilayered manganites. The reduction in the values of T${_N}$ with increasing substitution was observed to be similar to that observed in a well established 2D Heisenberg system  La${_2}$Cu${_{1-x}}$Mg${_x}$O${_4}$. This observation clearly indicates that the magnetic interactions in these bilayered systems are of the 2D Heisenberg type. The magnetic contribution to the specific heat near the transition temperature estimated using Fisher's relation could be fitted to a term corresponding to a gapped Fermi surface, the magnitude of which was seen to decrease with increasing substitution. The resistivity of the samples were seen to increase in the region T$<$T${_N}$ with increasing Al substitution due to the reduction of the effective double exchange interaction within each layer of the A type AFM, whereas in the paramagnetic region the conduction was seen to be of a simple 2D VRH form without a Coulomb gap.

\begin{figure}
\caption{The real part of first order ac susceptibility plotted as a function of temperature for the series  La${_{1}}$Sr${_{2}}$Mn${_{2-x}}$Al${_x}$O${_7}$.  The transition temperatures are seen to shift to lower temperatures with increasing Al doping.}
\caption{The normalised Neel temperature [T/T${_N}$] plotted as a function of substitution  for various quasi-2D systems. The data for K${_2}$(Co${_{1-x}}$Mg${_x}$)F${_4}$ and La${_2}$(Cu${_{1-x}}$Mg${_x}$)O${_4}$ are from ref \cite{breed} and \cite{swc} respectively. The skewed lines are guides to the eye indicating the effect of site dilution on a Mean Field and a Ising square lattice system}
\caption{The magnetic contribution to the specific heat plotted as a function of temperature for the series La${_{1}}$Sr${_{2}}$Mn${_{2-x}}$Al${_x}$O${_7}$. The solid line indicates fits incorporating the presence of a gapped Fermi surface. The inset shows the normalised value of $\Delta$ plotted as a function of the dopant concentration.}
\caption{Semilog plot of the resistivity($\rho$) vs T${^{-1/3}}$ of  La${_{1}}$Sr${_{2}}$Mn${_{2-x}}$Al${_x}$O${_7}$ for T $>$ T${_N}$.  The lines are linear fits to the experimental data indicating that the 2D variable range hoppping mechanism is valid in the paramagnetic region.}
\caption{Semilog plot of the resistivity($\rho$) as a function of temperature for the series La${_{1}}$Sr${_{2}}$Mn${_{2-x}}$Al${_x}$O${_7}$ , clearly indicating the increase in resistivity as a function of Al doping. This arises due to the weakening of the double exchange within the ferromagnetic layers in these A type antiferromagnets.}
\end{figure}

\begin{thebibliography}{ }
\bibitem[1]{RP}S. N. Ruddlesden and P. Popper, Acta Crystallogr.{\bf11}, 541 (1958).
\bibitem[2]{mori}Y. Moritomo, A. Asamitsu H. Kuwahara and Y. Tokura, Nature(London){\bf380}, 141 (1996).
\bibitem[3]{rao}Colossal Magnetoresistance, Charge Ordering and related properties of Manganese Oxides, ed. by C. N. R. Rao and B. Raveau, World Scientific, Singapore (1998).
\bibitem[4]{good1}J. B. Goodenough, Phys. Rev. {\bf121}, 229 (2002).
\bibitem[5]{tokura1}J. Q. Li, Y. Matsui, T. Kimura and Y. Tokura, Phys. Rev. B {\bf57}, R3205 (1998).
\bibitem[6]{chat}Tapan Chatterji, G. J. Mc'Intyre, W. Caliebe, R. Suryanarayanan, G. Dhalenne and R. Revcolevschi, Phys. Rev. B {\bf61}, 570 (2000). 
\bibitem[7]{kubota}M. Kubota, H. Yoshizawa, Y. Moritomo, H. Fujioka, K. Hirota and Y. Endoh, J. Phys. Soc. Jpn. {\bf68}, 2202 (1999).
\bibitem[8]{hardy}V. Hardy, A. Maignan, S. Hebert and C. Martin, Phys. Rev. B {\bf67}, 24401 (2003); A. Maignan, V. Hardy, C. Martin, S. Hebert and B. Reveau, J. Appl. Phys. {\bf93}, 7361 (2003).
\bibitem[9]{shan}R. D. Shannon and C. T. Prewitt, Acta. Cryst. {\bf{B25}}, 925 (1969).
\bibitem[10]{young}R. A. Young, A. Sakthivel, T. S. Moss, C. O. Paiva-Santos, \textit{Users Guide to Program DBWS-9411}, Georgia Institute of Technology, Atlanta, (1994).
\bibitem[11]{intergrowth}S. D. Bader, R. M. Osgood III, D. J. Miller, J. F. Mitchell and J. S. Jiang, J. Appl. Phys. {\bf83}, 6385 (1998). 
\bibitem[12]{ashna}A. Bajpai and A. Banerjee, Rev. Sci. Instrum. {\bf68}, 4075 (1997).
\bibitem[13]{breed}D. J. Breed, K. Gilijamse, J. W. E. Sterkenburg and A. R. Miedema, J. Appl. Phys. {\bf41}, 1267 (1970).
\bibitem[14]{swc}S-W. Cheong, A. S. Cooper, L. W. Rupp, Jr., B. Batlogg, J. D. Thompson and Z. Fisk, Phys. Rev. B {\bf44}, 9739 (1991).
\bibitem[15]{fisher}M. E. Fisher, Phil. Mag. {\bf7}, 1731 (1962).
\bibitem[16]{grif}J. Skalyo Jr., A. F. Cohen, S. A. Friedberg and R. B. Griffiths, Phys. Rev. {\bf164}, 705 (1967).
\bibitem[17]{maple}M. B. Maple, J. W. Chen, Y. Dalichaouch, T. Kohara, C. Rossel, M. S. Torikachvili, M. W. McElfresh and J. D. Thompson, Phys. Rev. Lett. {\bf56}, 185 (1986).
\bibitem[18]{esr}E. S. R. Gopal, \textit{Specific Heats at low temperatures} (Plenum Press, New York, 1966).
\bibitem[19]{sal}M. B. Salamon and M. Jaime, Rev. Mod. Phys. {\bf73}, 583 (2001).
\bibitem[20]{zhou}J. S. Jhou, J. B. Goodenough and J. F. Mitchell, Phys. Rev. B {\bf58}, R579 (1998).
\bibitem[21]{vrh}X. J. Chen, C. L. Zhang, J. S. Gardner, J. L. Sarrao and C. C. Almasan,Phys. Rev. B {\bf68}, 644051 (2003).
\bibitem[22]{mott}N. F. Mott and E. A. Davies, \textit{Electronic processes in Noncrystalline Solids}, 2nd ed, Clarendon Press, Oxford. (1979).
\bibitem[23]{kawano}H. Kawano, R. Kajimoto, H. Yoshizawa, Y. Tomioka, H. Kuwahara and Y. Tokura, Phys, Rev. Lett. {\bf78}, 4253 (1997).
\bibitem[24]{hiro}K. Hirota, Y. Moritomo, H. Fujioka, M. Kubota, H. Yoshizawa and Y. Endoh, J. Phys. Soc. Jpn. {\bf67}, 3380 (1998).
\bibitem[25]{mura}Y. Murakami, H. Kawada, M. Tanaka, T. Arima, Y. Moritomo and Y. Tokura,  Phys, Rev. Lett. {\bf80}, 1932 (1998).
\bibitem[26]{sunil}Sunil Nair and A. Banerjee, cond-mat/0309406.
\end{thebibliography}
\end{document}